\documentclass[11pt,a4paper]{article}
\usepackage[top=1in, bottom=1in, left=1in, right=1in]{geometry}

\usepackage{amsmath}
\usepackage{siunitx}        
\usepackage[xindy,nomain,acronym]{glossaries}
\usepackage{amssymb}
\usepackage{graphicx}
\usepackage{subcaption}

\usepackage[backend=bibtex,natbib=true, maxcitenames=1]{biblatex} 
\usepackage{color, colortbl}

\title{Fr\'echet Audio Distance: A Metric for Evaluating Music Enhancement Algorithms}
\author{Kevin Kilgour, Mauricio Zuluaga, Dominik Roblek, Matthew Sharifi \\ Google AI}
\date{}                     

\newacronym{SDR}{SDR}{signal to distortion ratio}
\newacronym{SNR}{SNR}{signal to noise ratio}
\newacronym{SIR}{SIR}{signal to interference ratio}
\newacronym{FAD}{FAD}{Fr\'echet Audio Distance}
\newacronym{FID}{FID}{Fr\'echet Inception Distance}
\newacronym{unet}{U-net}{U-net, named after its U-shaped structure}
\newacronym{STOI}{STOI}{short-time objective intelligibility}
\newacronym{PESQ}{PESQ}{Perceptual evaluation of speech quality}
\newacronym{dwn}{discriminative Wavenet}{discriminative Wavenet}
\newacronym{munet}{MU-net}{modified U-net}
\newacronym{ASR}{ASR}{automatic speech recognition}
\newacronym{WER}{WER}{word error rate}
\begin{document}
%

\maketitle

\begin{abstract}
We propose the \gls{FAD}, a novel, reference-free evaluation metric for music enhancement algorithms. We demonstrate how typical evaluation metrics for speech enhancement and blind source separation can fail to accurately measure the perceived effect of a wide variety of distortions. As an alternative, we propose adapting the \gls{FID} metric used to evaluate generative image models to the audio domain. \gls{FAD} is validated using a wide variety of artificial distortions and is compared to the signal based metrics \gls{SDR}, cosine distance and magnitude L2 distance. We show that, with a correlation coefficient of $0.52$, \gls{FAD} correlates more closely with human perception than either \gls{SDR}, cosine distance or magnitude L2 distance, with correlation coefficients of $0.39$, $-0.15$ and $-0.01$ respectively. 
\end{abstract}

\glsresetall

\section{Introduction}
\label{sec:intro}
Music enhancement aims to accomplish two goals: separating a music signal from other, interfering noise signals and improving its quality to sound more like studio recorded music. Imagine a mobile phone recording of Vivaldi's Four Seasons played through low quality speakers in a large, reverberant room where a group of people are having a loud discussion. The resulting recording will not be pleasant to listen to.

Video hosting platforms such as YouTube \cite{youtube} and Vimeo \cite{vimeo} contain a multitude of amateur musical recordings, often captured with a low quality microphone in a setup very different from a recording studio. Such recordings could potentially benefit from music enhancement.

Existing research has looked into techniques for speech separation \cite{wang2017supervised} and speech enhancement \cite{Loizou:2013:SET:2484638} as well as separating music into its instrumental components \cite{park2018music} or removing the vocals to produce a karaoke version of the track \cite{jansson2017singing}. Speech enhancement and separation have been active areas of research for many years. Applications include enhancing mobile device recordings, hearing aids and conference call systems.

For the specific task of music enhancement, we found it challenging to quantitatively compare different approaches or models with respect to the perceived quality of their output.

Standard metrics\footnote{Throughout this paper, the term metric will be used to mean a measure for quantitative assessment and not necessarily a mathematical measure of distance.} such as \gls{SDR} and \gls{SIR} \cite{vincent2006performance}, which are typically used to evaluate signal separation algorithms, are able to determine which music enhancement algorithm produces reconstructed music whose signal is closest to a studio recorded original. However, these metrics do not take into account the perceptual quality of the reconstructed music which sometimes results in reconstructions with a lower \gls{SDR} being more pleasing to listen to. A further disadvantage is that these metrics are full-reference metrics and require a copy of the studio recorded music that the enhancement algorithm should produce.

Based on the \gls{FID}, introduced by \citet{heusel2017gans} to evaluate generative models for images, we propose the \gls{FAD} for evaluating generated audio. \gls{FAD} compares statistics computed on a set of reconstructed music clips to background statistics computed on a large set of studio recorded music. We compare both \gls{SDR} and \gls{FAD} against human ratings to evaluate their correlation with perceptual quality.

\section{Related Work}
\label{sec:related}
In speech enhancement there are three overarching approaches for evaluating the quality of the speech: direct signal comparison methods, human evaluations, and signal-based heuristics which are designed to correlate with human evaluation scores.

The first type of approach compares the enhanced speech signal to a reference signal. This includes basic distance metrics such as \textit{cosine distance} and \textit{$L_2$ distance} as well as ratio metrics such
as \gls{SNR}, \gls{SDR} and \gls{SIR} \cite{vincent2006performance}. These full-reference  metrics are agnostic to the type of audio being separated or enhanced and can be used for evaluating music enhancement techniques without any changes.

Throughout this paper, we use the implementation of \gls{SDR} from \citet{raffel2014mir_eval}. \citet{roux2018sdr} have recently brought to light some weaknesses of \gls{SDR} and this implementation in particular. They propose a \textit{scale invariant \gls{SDR}} as an alternative.

Although useful, signal level metrics do not necessarily predict how a human listener will perceive the reconstructed music. For speech enhancement, perceptual level metrics are regularly used, where human raters are asked to compare speech output with ground truth. Human raters are typically provided with individual audio clips of speech and asked to evaluate the \textit{naturalness} of the speech signal on a five point scale from \textit{5 (very natural, no degradation)} down to \textit{1 (very unnatural, very degraded)} and how intrusive the background noise is from \textit{5 (not noticeable)} down to \textit{1 (very conspicuous, very intrusive)} \cite{hu2008evaluation}. 

The third category of speech enhancement metrics, which are not trivially applicable to evaluating music enhancement approaches, are automatic metrics such as \gls{PESQ} \cite{rix2001perceptual} and \gls{STOI} \cite{taal2010short} that approximate perceptual level metrics without requiring any human raters. Such metrics are designed to correlate with human evaluation scores for speech quality specifically.

When the result of a speech enhancement algorithm is primarily intended to be used as the input to a subsequent process, e.g. an \gls{ASR} system, then it makes sense to measure its quality using an error metric designed for that subsequent process. \citet{2018arXiv181111517C} developed a method called acoustics-guided evaluation which uses an existing acoustic model and compares the posteriors of an enhanced evaluation set to the posteriors of its aligned, cleaned counterpart. The authors showed that this metric is highly correlated with \gls{WER}.

In this paper, we propose an automatic metric designed for music enhancement, which is based on the \gls{FID} metric used to evaluate image-generating GANs. \gls{FID} uses the coding layer of the Inception network \cite{szegedy2015going} to generate embeddings from an evaluation set of images produced by the GAN and a large set of background images. The Fr\'echet distance \cite{dowson1982frechet} is then computed between multivariate Gaussians estimated on the evaluation embeddings and the background embeddings. This approach has also been adapted to videos by \citet{unterthiner2018towards}.

\section{Fr\'echet Audio Distance (FAD)}

Through our initial work developing techniques for music enhancement, we observed that signal based metrics often disagreed with our own subjective evaluations of the enhanced music. These metrics would penalize enhanced music that differed from the ground truth signal, even when it would sound more like studio quality music to a human listener. To this end, we propose \gls{FAD}: a metric which is designed to measure how a given audio clip compares to clean, studio recorded music.

\begin{figure*}[t]
\includegraphics[width=0.98\linewidth]{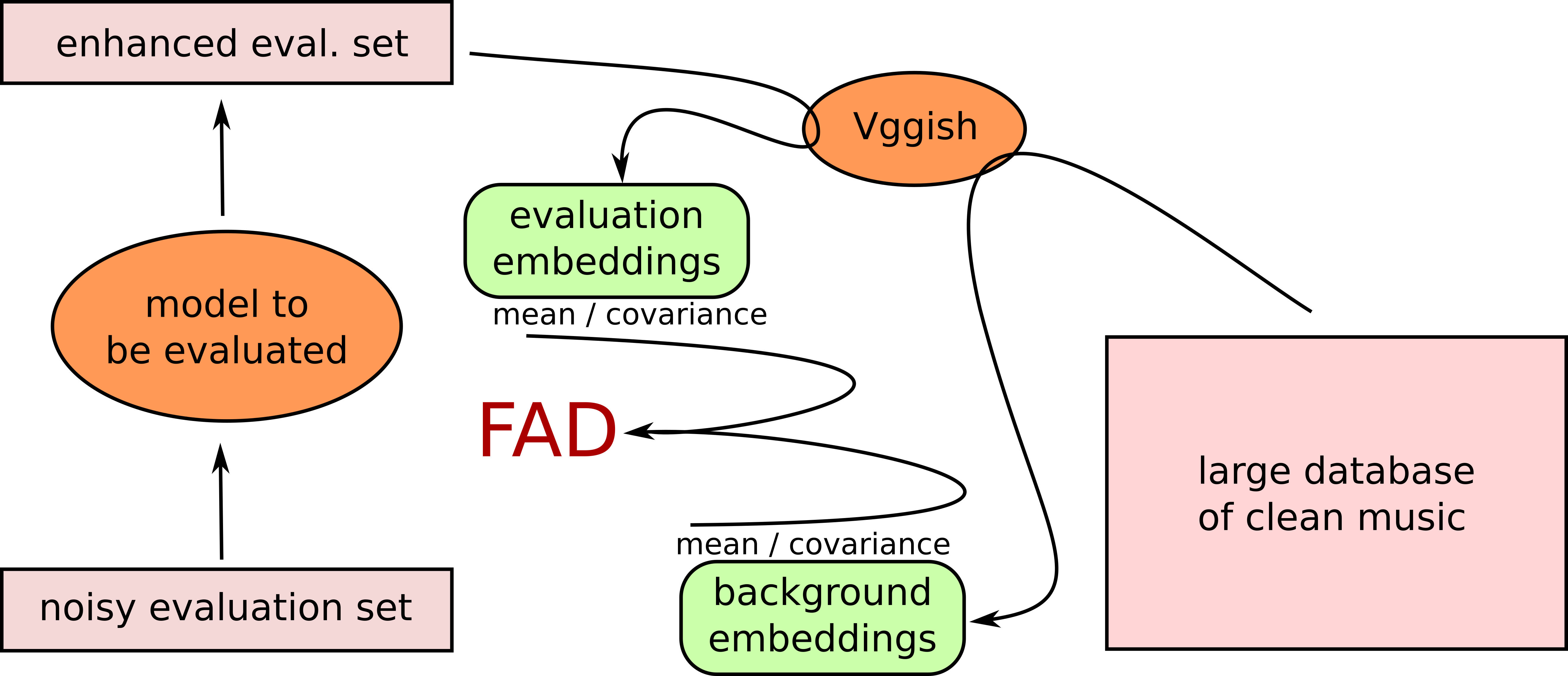}
\caption[FAD computation overview]{\textit{\gls{FAD} computation overview: using a pretrained audio classification model, VGGish, embeddings are extracted from both the output of a enhancement model that we wish to evaluate and a large database of background music. The Fr\'echet distance is then computed between multivariate Gaussians estimated on these embeddings.}}
\label{fig:overview}
\end{figure*}

\subsection{Definition}

Unlike existing audio evaluation metrics, \gls{FAD} does not look at individual audio clips, but instead compares embedding statistics generated on the whole evaluation set with embedding statistics generated on a large set of clean music (e.g. the training set).
This makes \gls{FAD} a reference-free free metric which can be used to score an an evaluation set where the ground truth reference audio is not available. Where \gls{FID} uses the activations from a hidden layer in the Inception network \cite{szegedy2015going} to generate embeddings, \gls{FAD} uses embeddings generated by the VGGish \cite{hershey2017cnn} model.

As shown in Figure \ref{fig:overview}, this gives us a set of background embeddings from the clean music and a set of evaluation embeddings from the output of the music enhancement model that we wish to evaluate. 

We then compute multivariate Gaussians on both the evaluation set embeddings $\mathcal{N}_e(\mu_e,\,\Sigma_e)$ and the background embeddings $\mathcal{N}_b(\mu_b,\,\Sigma_b)$. \citet{dowson1982frechet} show that the Fr\'echet distance between two Gaussians is:
\begin{align}
    \textbf{F}(\mathcal{N}_b,\mathcal{N}_e) = \| \mu_b - \mu_e \|^2 +tr(\Sigma_b + \Sigma_e - 2 \sqrt{\Sigma_b \Sigma_e})
\end{align}
where $tr$ is the trace of a matrix. When comparing models, both the background embeddings and the evaluation set of noisy signals passed as input to the model are fixed. We often refer to the \gls{FAD} computed between embeddings of the denoised evaluation set and the background embeddings as a model's \textit{\gls{FAD} score}.

\subsection{FAD Embedding Model}
 VGGish\footnote{VGGish can be downloaded from: \\https://github.com/tensorflow/models/tree/master/research/audioset} is derived from the VGG image recognition architecture \cite{DBLP:journals/corr/SimonyanZ14a} and is trained on a large dataset of YouTube videos, similar to YouTube-8M \cite{abu2016youtube} as an audio classifier with over \num{3000} classes. The activations from the 128 dimensional layer prior to the final classification layer are used as the embedding.

The input to the VGGish model consists of 96 consecutive frames of 64 dimensional log-mel features extracted from the magnitude spectrogram computed over \SI{1}{s} of audio. Given that the input requirement of \SI{1}{s} is considerably shorter than typical evaluation music clips, we extract \SI{1}{s} windows every \textit{t} seconds. In Appendix~\ref{app:window_step}, we analyze what value should be chosen for \textit{t} and find that it should be \SI{0.5}{s}, thereby overlapping each window by \SI{50}{\%}.

It is worth noting that the input to the existing VGGish model may not be ideal, given that ignoring the phase and using mel-scaled bins could lead to certain distortions going undetected. We investigate this further in Section~\ref{sec:mel}.

\section{Experimental Setup}
To verify the usefulness of our FAD metric, we start by firstly computing the background statistics $\mathcal{N}_b$ over embeddings from a dataset of clean music. We then apply various distortions to our audio clips from the evaluation set and compute statistics on their embeddings. The distortions can be viewed as both artifacts that could possibly be introduced by a music enhancement algorithm, or as interfering noises that were not completely removed. We obtain an \gls{FAD} score for each parameter configuration of a distortion. 

\subsection{Artificial Distortions}
\label{sec:dist}

The intensity of each distortion can be controlled by one or more parameters. We expect that, for a given distortion function, parameter configurations which distort the audio more should have a higher \gls{FAD} score.

\begin{description}
    \item[Gaussian noise]: A distortion signal is sampled from a normal distribution, with $\mu = 0$ and varying $\sigma$, and added to the input signal.
    \item[Pops]: We randomly select p\% of the input signal's samples and set half of them to $-1$ and half to $+1$, or $\pm \texttt{abs}(\texttt{max\_input})$ if the  signal is not normalized.
    \item[Frequency filter]: The signal is passed through a high or low pass filter with various cutoff frequencies. 
    \item[Quantization]: The signal is reduced from \SI{16}{bits} per sample down to $q$~bits per sample.  
    \item[Griffin-Lim distortions]: The signal is converted to a magnitude spectrogram, and then reconstructed using the Griffin-Lim algorithm \cite{1164317}. The quality of the reconstructed phase depends on the algorithm's iteration parameter.
\item[Mel encoding]: The signal is converted into a mel-scale magnitude spectrogram and back again using the original input phase. We look at two mel variants: narrow, where the mel bins only covers the frequency range from 60 to \num{6000}~Hz, and wide, which covers everything from 0 to \num{16000}~Hz. 
    \item[Speed up / slow down]: The playback speed of the signal is increased/decreased by a given factor, and as a side effect this also leads to an increase/decrease in its pitch.
    \item[Pitch preserving speed up / slow down]: The playback speed of the signal is increased/decreased by a given factor using a \textit{phase vocoder} \cite{flanagan1966phase} which preserves the signal's original pitch.
    \item[Reverberations]: Multiple dampened copies of the original signal are added using a provided delay.
    \item[Pitch up / down]: The pitch of the signal is increased/decreased by a provided number of semitones.
\end{description}
All distortions are designed to be unaffected by loudness normalization. The distortions for each test parameter configuration are applied separately and in parallel to each of the audio segments in the evaluation set to generate embeddings. This results in an \gls{FAD} score for each distortion parameter configuration. An overview of the parameters used for each distortion can be found in Appendix~\ref{app:params}.

\subsection{Data}
\label{sec:data}
For our experiments, we use the Magnatagatune dataset \cite{Law_evaluationof}, which contains \SI{600}{hours} of music samples at \SI{16}{kHz}. We use \SI{540}{hours} as the background clean music set and \SI{60}{hours} for evaluation of the metrics. For human evaluations, a \SI{25}{minute} subset of the \SI{60}{hour} evaluation set is used, which is split into $300$ audio clips of \SI{5}{s} in length.

\subsection{Evaluation Metrics}
\label{lab:data}
In addition to \gls{FAD}, we compute the cosine distance, magnitude L2 distance and \gls{SDR} scores of each parameter configuration of the distortions using:
\begin{align}
\texttt{SDR}(s_d, s_c) &= \frac{\|s_\texttt{target}\|^2}{\|e_\texttt{interf} + e_\texttt{noise} + e_\texttt{artif}\|^2} \\
\texttt{magnitudeL2}(s_d, s_c) &= \||\texttt{stft}(s_d)|-|\texttt{stft}(s_c)|\|_{2} \\
\texttt{cosdist}(s_d, s_c) &= 1 - \texttt{cosim}(s_d, s_c) = \frac{s_d \cdot s_c}{\|s_d\| \|s_c\|}
\end{align}
where $s_d$ is the distorted audio signal, $s_c$ the corresponding clean audio signal. Please refer to \citet{vincent2006performance} for more details on \gls{SDR}.

The output range of cosine distance is between $0$ and $2$, where values closer to $0$ indicate that the signals are more positively correlated, values closer to $2$ that they are more negatively correlated, and values close to $1$ that they are either not at all, or only insignificantly correlated. This follows from the definition of the cosine similarity. We omit the $L_2$ distance on samples from our analysis because, for normalized signals, a target $t$, and output $o$, the $L_2$ distance $\|t - o\|_2$ is $\sqrt{2~\texttt{cosdist}(t,o)}$, which does not provide us with any further information relevant to the evaluation. Unlike the other metrics where lower values are better, \gls{SDR} scores signals that are more similar higher. As a result, we plot -SDR to maintain a consistent pattern of lower being considered as better.

\begin{figure}[t]
    \begin{subfigure}[b]{1\textwidth}
      \includegraphics[width=0.98\linewidth]{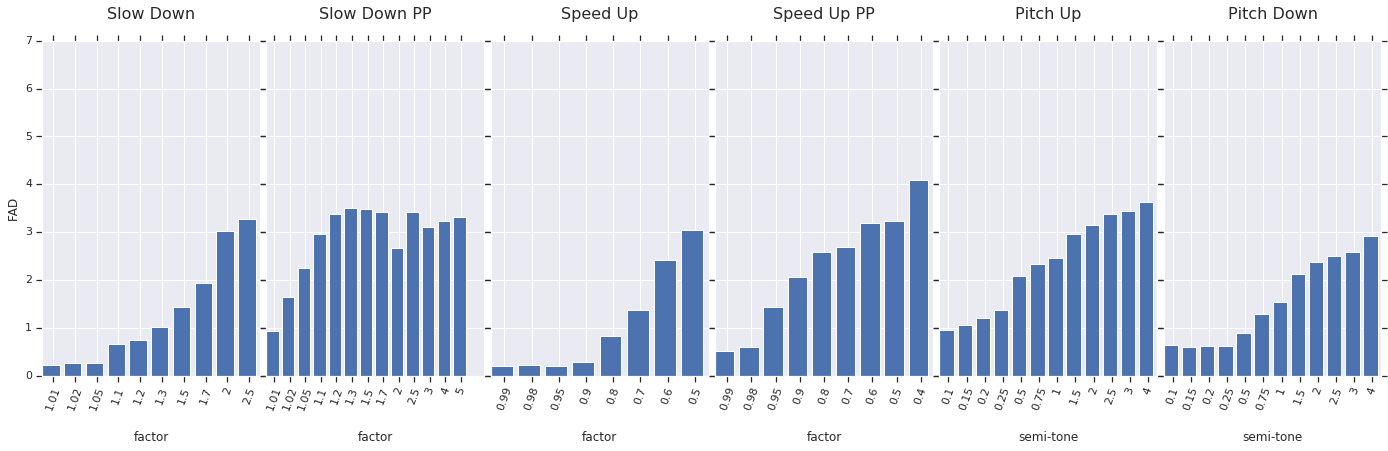}
    \end{subfigure}
    
    \begin{subfigure}[b]{1\textwidth}
      \includegraphics[width=0.98\linewidth]{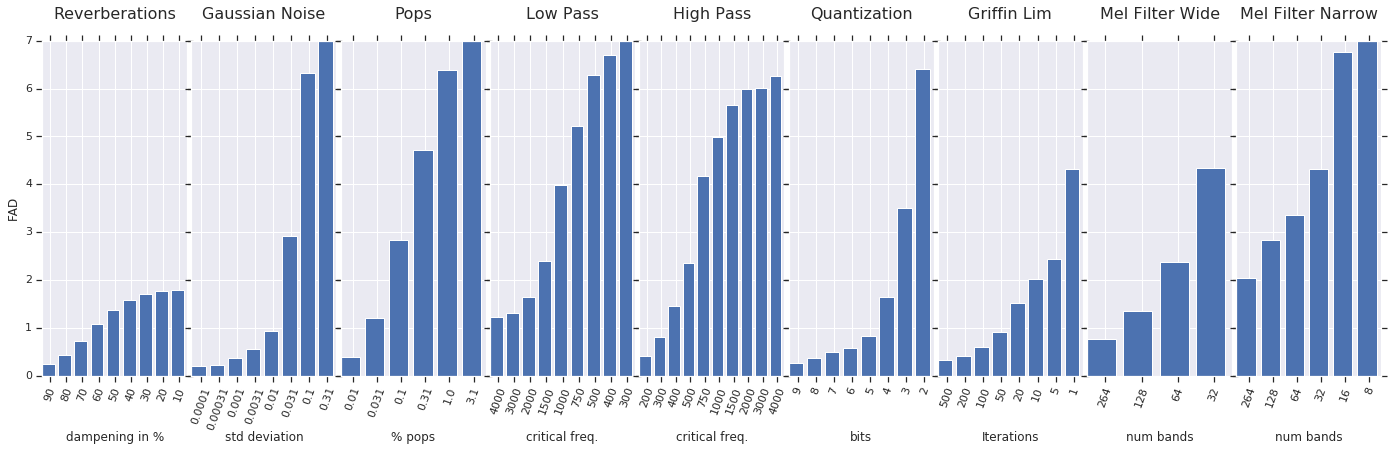}
    \end{subfigure}
\caption[FAD scores]{\textit{FAD scores for artificial distortions on the Magnatagatune dataset. The abbreviation PP indicates the pitch preserving variant of a distortion. For comparison, the FAD computed on the non-distorted clean audio is $0.2$.}}
\label{fig:dist}
\end{figure}

\subsection{Human Evaluation}
For our human-based evaluation, we asked raters to compare the effect of two different distortions on the same \SI{5}{s} of audio, randomizing both the pair of distortions that they compared and the order in which they appeared. We included the clean original as a pseudo-distortion. The raters were asked \textit{``which audio clip sounds most like a studio produced recording?''} and if they were unable to make a choice after listening to both clips twice, they were able to declare them tied.

\begin{figure}[t]
\includegraphics[width=1\linewidth]{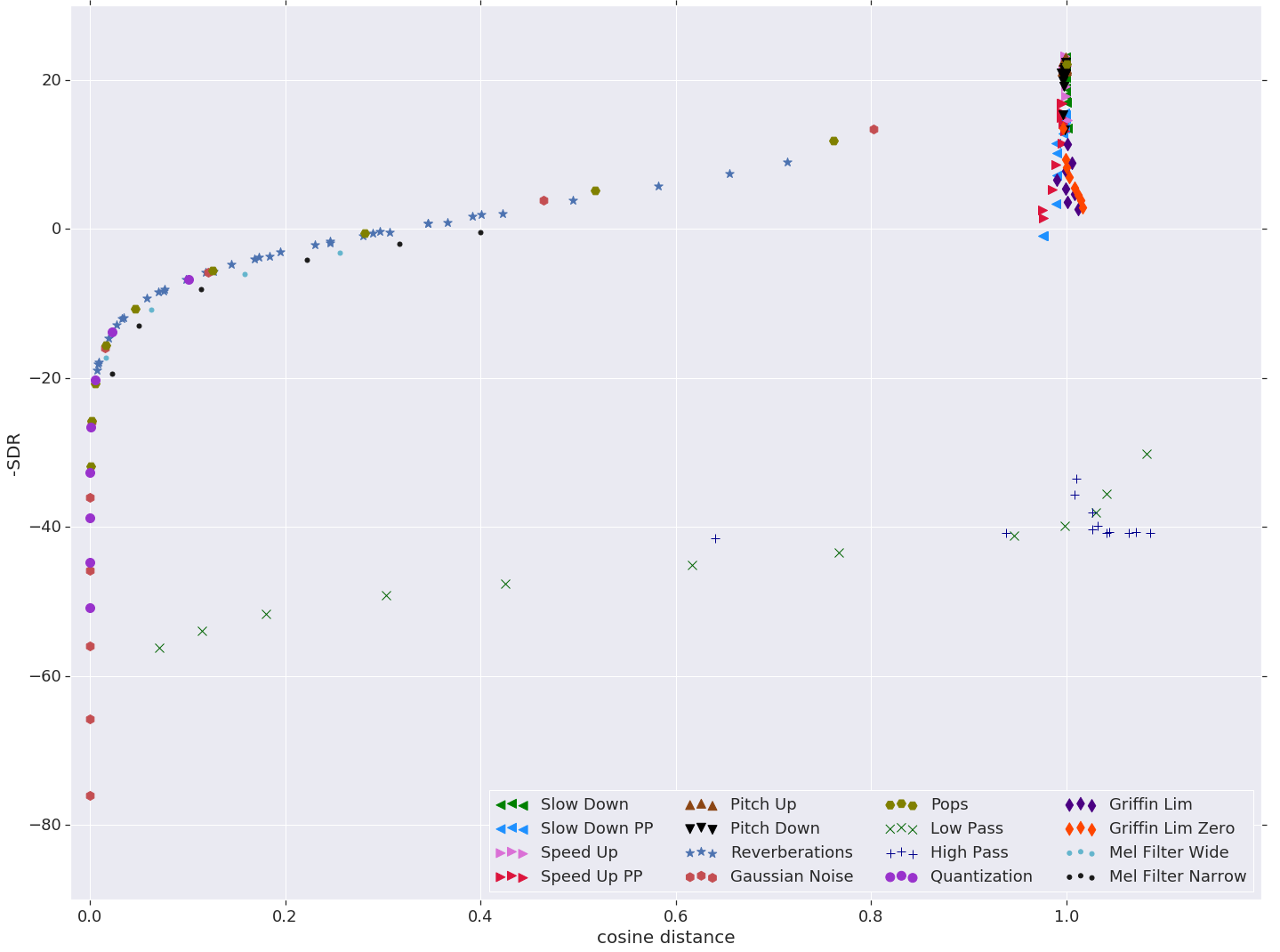}
\caption{\textit{Comparison of \gls{SDR} and cosine distance. The abbreviation PP indicates the pitch preserving variant of a distortion. A full overview of the parameter values tested for each distortion can be found in Appendix~\ref{app:params}. }}
\label{fig:SDR_cos}
\end{figure}

\begin{figure}[t]
\includegraphics[width=1\linewidth]{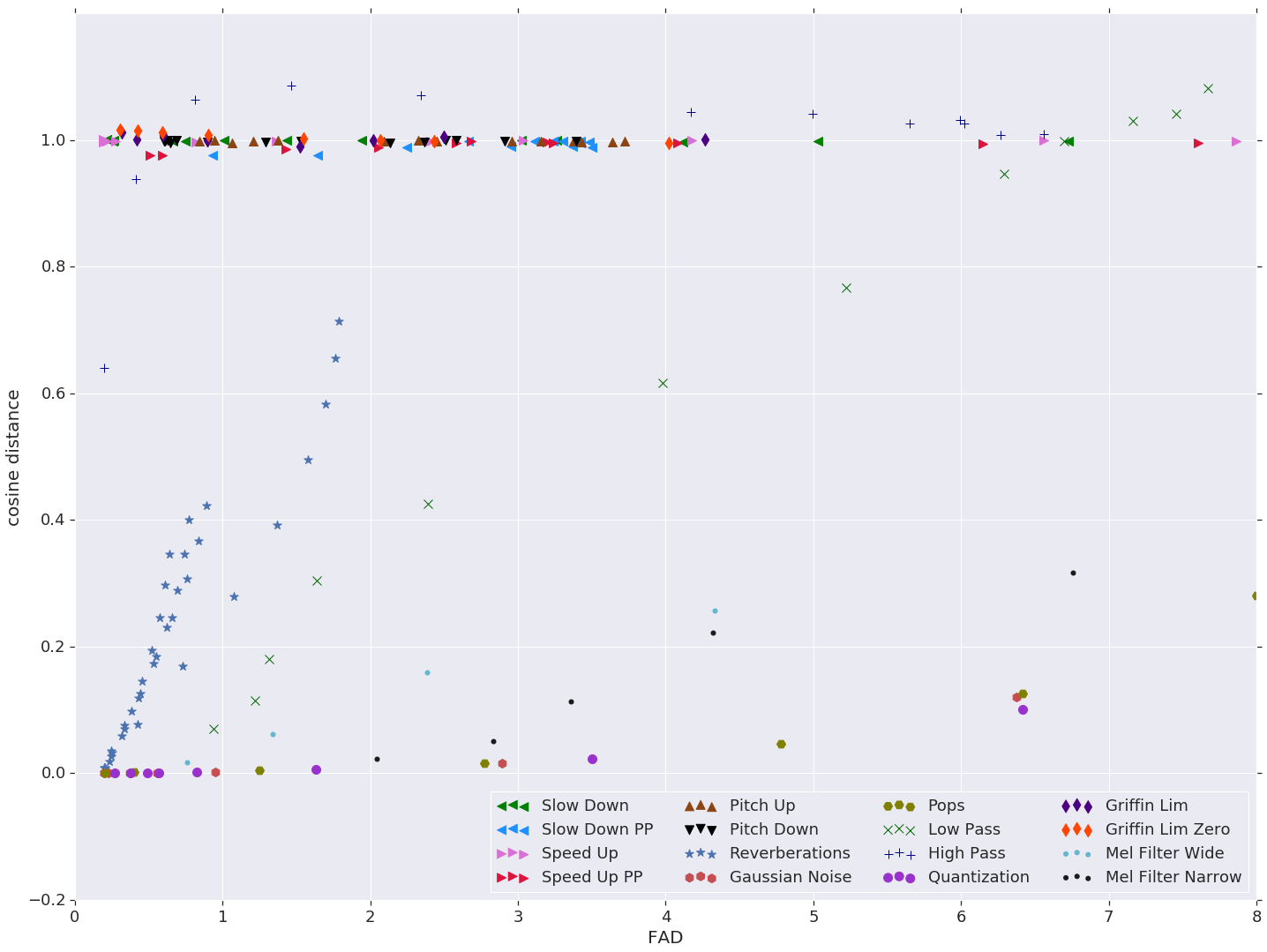}
\caption{\textit{Comparison of \gls{FAD} and cosine distance. The abbreviation PP indicates the pitch preserving variant of a distortion. A full overview of the parameter values tested for each distortion can be found in Appendix~\ref{app:params}. }}
\label{fig:fad_cos}
\end{figure}

\section{Results}
\label{sec:mel}
An overview of the \gls{FAD} scores on the distortions described in Section~\ref{sec:dist} with various parameters is shown in Figure~\ref{fig:dist}. Overall, the \gls{FAD} scores of the distortions generally behave as expected, with \gls{FAD} scores increasing as the magnitude of the distortion is increased. For the Gaussian noise distortion, the low \gls{FAD} scores for very small standard deviations are reasonable because such distortions are also barely detectable to a human. Their FAD scores of $0.2-0.3$ are almost the same as the FAD score of $0.2$ computed on non-distorted clean audio. 

We observed that distortions with similar \gls{FAD} scores were of similar subjective quality, e.g. we perceived Gaussian noise with a standard deviation of $0.031$ as having roughly the same quality as setting the percentage of pops to $0.1$, and slightly worse than quantizing to \SI{4}{bits}. In Section~\ref{sec:human} we run a large scale human evaluation in order to validate our subjective observations.

We verified that using an embedding model which only looks at a mel-scale magnitude spectrogram could still be useful in identifying phase distortions. Removing the phase and reconstructing the signal using Griffin-Lim is noticeable to humans, but often results in audio with an acceptable quality given a sufficient number of iterations. With an iteration parameter of $5$, the Griffin-Lim distortion had an \gls{FAD} score of $2.4$. This steadily decreased when the iteration parameter was increased, plateauing out at around $0.31$ after $500$ iterations. 

Applying a mel filter is also detectable using \gls{FAD}. A wide mel filter with \SI{64}{bins} results in an \gls{FAD} score of $2.4$, while using \SI{32}{bins} increases the \gls{FAD} score to $4.3$. Even using \SI{256}{bins} results in detectable FAD scores for both the narrow and wide variants. These last two results highlight the usefulness of \gls{FAD} in detecting distortions and irregularities in music signals, and indicate that it should prove useful in evaluating music enhancement models.

\subsection{Comparison to Signal Based Metrics}

\begin{figure}[t]
\includegraphics[width=1\linewidth]{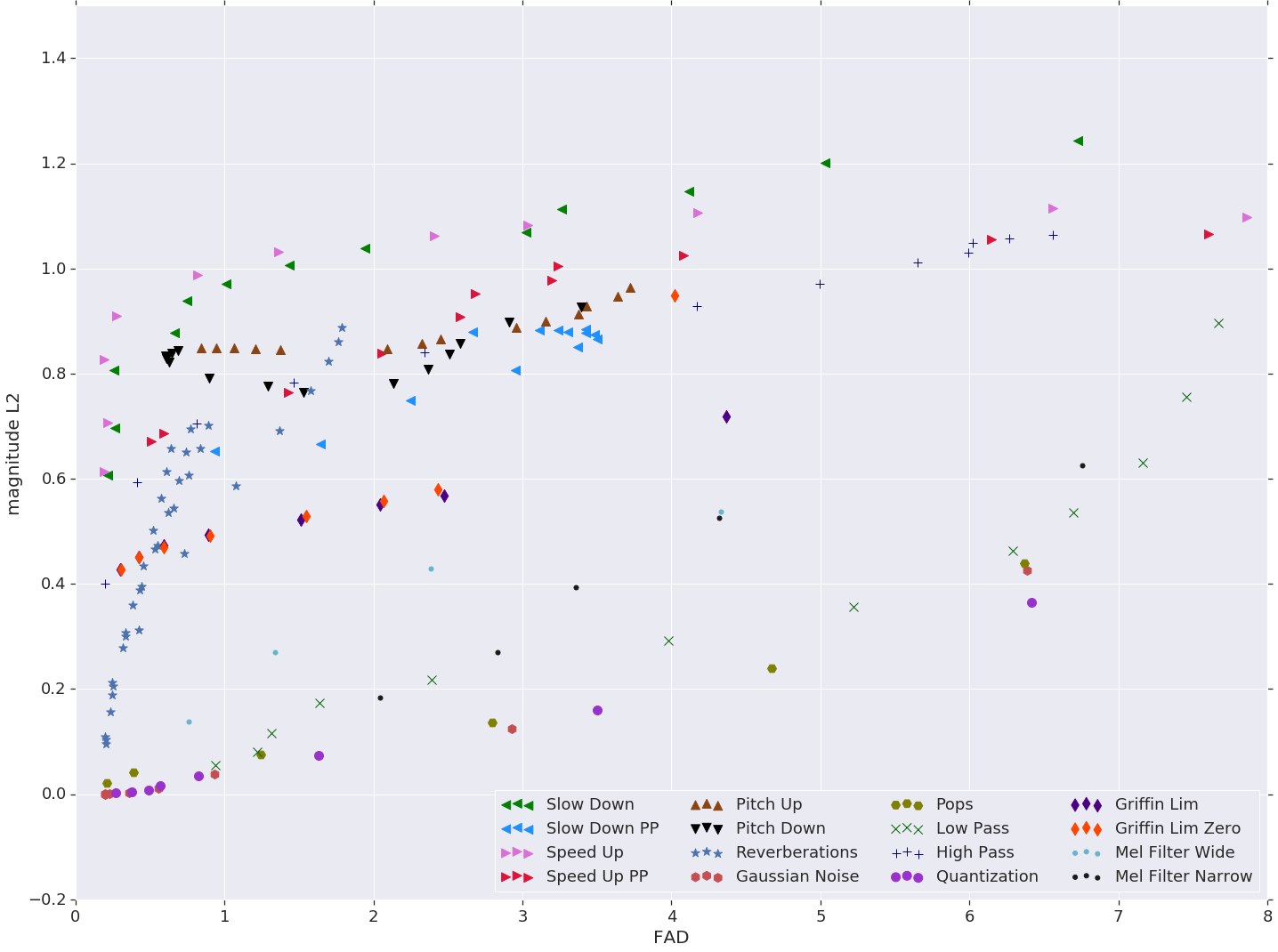}
\caption{\textit{Comparison of \gls{FAD} and magnitude L2 distance. The abbreviation PP indicates the pitch preserving variant of a distortion. A full overview of the parameter values tested for each distortion can be found in Appendix~\ref{app:params}.  }}
\label{fig:fad_magl2}
\end{figure}

\begin{figure}[t]
\includegraphics[width=1\linewidth]{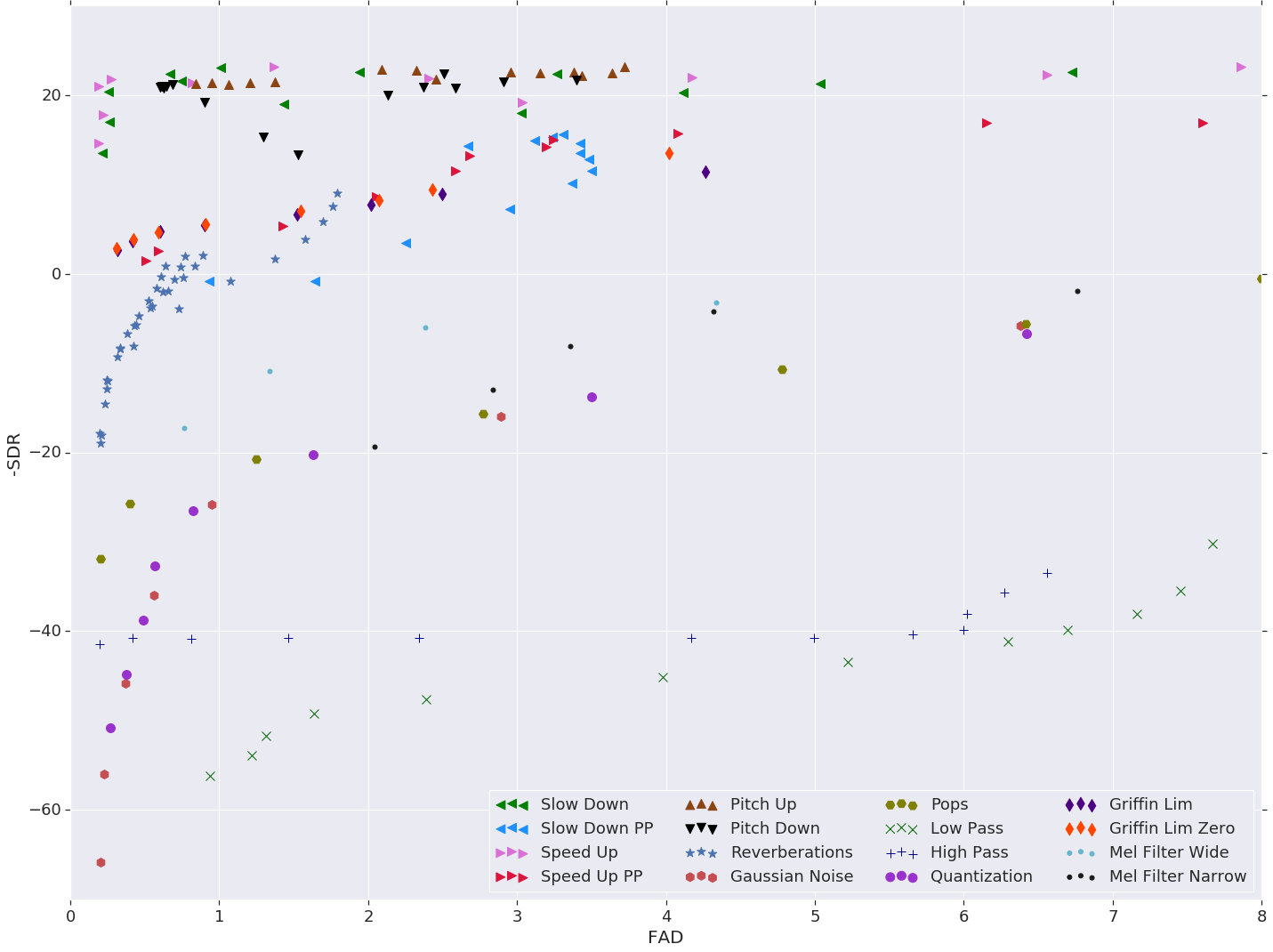}
\caption{\textit{Comparison of \gls{FAD} and \gls{SDR}. The abbreviation PP indicates the pitch preserving variant of a distortion. A full overview of the parameter values tested for each distortion can be found in Appendix~\ref{app:params}.  }}
\label{fig:fad_sdr}
\end{figure}

In this section, we compare how different distortions affect \gls{SDR}, \gls{FAD} and cosine distance. Figure~\ref{fig:SDR_cos} shows the \gls{SDR} of distortions at various parameter configurations and their cosine distance with three distinct groups of distortions clearly visible. The boomerang shaped group consists of \textit{gaussian noise}, \textit{quantization}, \textit{mel filter}, \textit{pops} and \textit{reverberations} which mostly function as additive distortions and do not affect the signal temporally. For these distortions, their \gls{SDR} score is proportional to the logarithm of their cosine distance. 

The second group, forming a narrow band in top right corner, consists of \textit{speed up/slow down}, \textit{pitch preserving speed up/slow down}, \textit{pitch up/down} and \textit{Griffim-Lim} which displace the signal from its reference by either stretching/compressing the signal or by altering its phase. Each distortion in this group has a cosine distance value of $1$, indicating that the signals are completely different as far as cosine distance is concerned. These distortions also result in variable and generally low \gls{SDR} scores. Because \gls{SDR} allows for time-invariant filter distortions up to a fixed number of samples $T$, it can still catch differences between them up to a certain extent.

The final group containing only the high and low pass filters have a cosine distances that are to be expected but surprisingly high \gls{SDR} scores. This is again due to \gls{SDR} being insensitive to certain transformations, which is explored in detail by \citet{roux2018sdr}.

Comparing \gls{FAD} to cosine distance (Figure~\ref{fig:fad_cos}), we again see two distinct groups of distortions. Along the top of the figure, we find \textit{speed up/slow down}, \textit{pitch preserving speed/slow down}, \textit{pitch up/down}, \textit{Griffin-Lim} and \textit{high pass} all have cosine distance values of $1$, regardless of the amount of distortion applied. On the other hand, their \gls{FAD} values are almost always monotonic and increase when the level of distortion of is increased. The other distortions, \textit{gaussian noise}, \textit{quantization}, \textit{mel filter}, \textit{pops}, \textit{high pass} and \textit{reverberations} appear to be correlated on an individual basis but not between distortions. This implies that, while both metrics can detect these distortions, they rate their severity differently. The cosine distance penalizes \textit{reverberations} and \textit{high pass} more than \gls{FAD} which is more affected by \textit{gaussian noise}, \textit{quantization}, \textit{mel filter} and \textit{pops}.

In Figure~\ref{fig:fad_magl2}, we show \gls{FAD} plotted against magnitude L2 distance. Overall the distortions appear to be individually correlated on log scale, but the two metrics disagree a lot regarding how intense the distortions are. The distortions \textit{gaussian noise}, \textit{quantization}, \textit{mel filter}, \textit{high pass} and \textit{pops} are more highly penalized by \gls{FAD}, while the others, with the exception of \textit{Griffin-Lim}, are much more highly penalized by magnitude L2 distance. All parameter configurations of the \textit{Griffin-Lim} distortion have magnitude L2 distances that only vary between $0.4$ and $0.6$ while their \gls{FAD} scores are more spread out.

The \gls{FAD} to \gls{SDR} plot in Figure~\ref{fig:fad_sdr} is more spread out. As before, we see that \gls{SDR} is almost invariant to the \textit{high pass} and \textit{low pass} distortions. Because \gls{FAD} does a very good job of detecting these distortions, they form a band along the bottom of the plot. Another band containing \textit{speed up}, \textit{slow down} and \textit{pitch up/down} along the top of the plot are the distortions that consistently get a low \gls{SDR} score regardless of their intensity, while \gls{FAD} increases with an increase in intensity.

For the remaining distortions, we see that each distortion's log \gls{FAD} scores are correlated with its \gls{SDR} scores. We observe that the two metrics rate the distortion types differently, with \gls{FAD} again penalizing \textit{gaussian noise}, \textit{quantization}, \textit{mel filter} and \textit{pops}. \gls{SDR} on the other hand is more tolerant of them and gives \textit{reverberations}, \textit{Griffin-Lim}, and \textit{pitch preserving speed up/slow down} high scores.

\subsection{Analysis}

Taken as a whole, these comparison plots split the distortions into 4 groups:
\begin{description}
    \item[\gls{SDR}-breaking distortions]: These are distortions that will lead to very low \gls{SDR} scores, independent of the distortion parameter configuration. They will generally have high magnitude L2 distances and their cosine distance score will be around $1$. Their  \gls{FAD}, on the other hand, appears to be sensitive to these distortions, with a low intensity of distortion having low FAD scores which increases for parameter configurations that cause more intense distortions. For some distortions, \gls{FAD} appears to plateau by either always having a minimum value no matter how low the distortion parameter, or no longer increasing after a certain maximum distortion parameter.
    The group includes: \textit{speed up}, \textit{slow down} and \textit{pitch up/down}  
    
    \item[Somewhat \gls{SDR}-breaking distortions]: The distortions in this group have low \gls{SDR} scores that vary with their distortion parameter configuration. While \gls{SDR} can also detect the intensity of the distortion, their scores will still be very low even for low distortion levels. Their cosine distance is either continuously $1$, or behaves similarly to \gls{SDR}. \gls{FAD} treats these distortions the same as the \gls{SDR}-breaking distortions. Their magnitude L2 distances are medium to high. They include: \textit{reverberations}, \textit{Griffin-Lim}, and \textit{pitch preserving speed up/slow down}.
    
    \item[Mainline distortions]: For these distortions, all four metrics are low for low parameter configurations and progressively increase as the amount of distortion increases, although for some of them \gls{FAD} may still plateau on both the low and high ends. Their rate of increase varies by distortion. This group includes: \textit{gaussian noise}, \textit{quantization}, \textit{mel filter} and \textit{pops}
    
    \item[\gls{SDR}-tolerant distortions]: Unlike \gls{FAD}, \gls{SDR} has a hard time detecting \textit{low pass} and \textit{high pass} filters. For the cosine distance \textit{low pass} behaves like the breaking distortions, while \textit{high pass} is detectable and behaves like the mainline distortions. According to the magnitude L2 distance \textit{high pass} behaves more like a somewhat \gls{SDR}-breaking distortion and \textit{low pass} like a mainline distortion.
    
\end{description}
\subsection{Human Evaluation}
\label{sec:human}

\begin{figure}[ht]\centering
\includegraphics[width=0.82\linewidth]{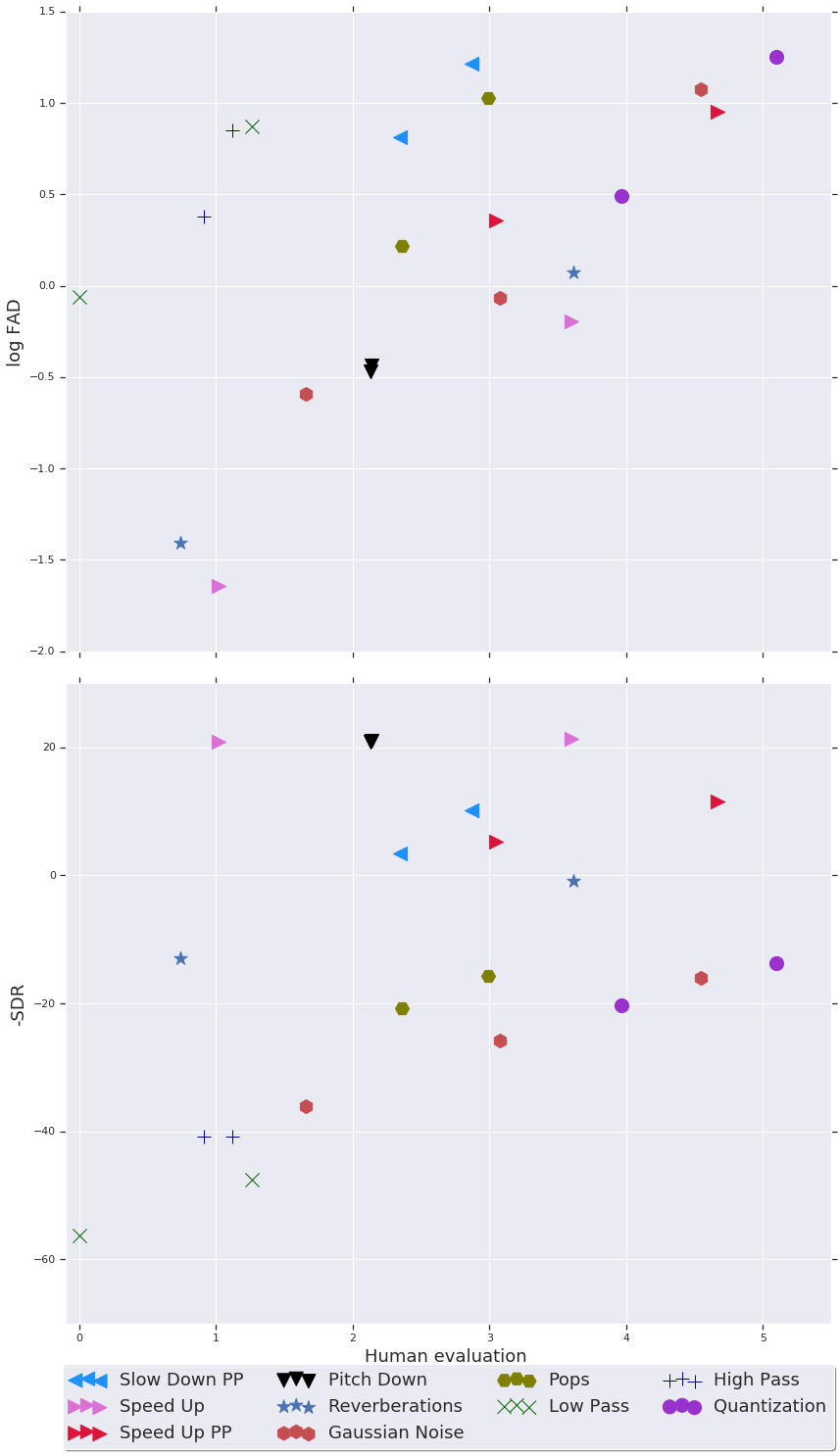}
\caption{\textit{Results of our human evaluation. The scale on x-axis is the worth value estimated by our Plackett-Luce model. The top plot compares this worth value to the distortions \gls{FAD} score and the bottom plot compares it to \gls{SDR}.}}
\label{fig:sdr_human}
\end{figure}

\begin{figure}[ht]\centering
\includegraphics[width=0.82\linewidth]{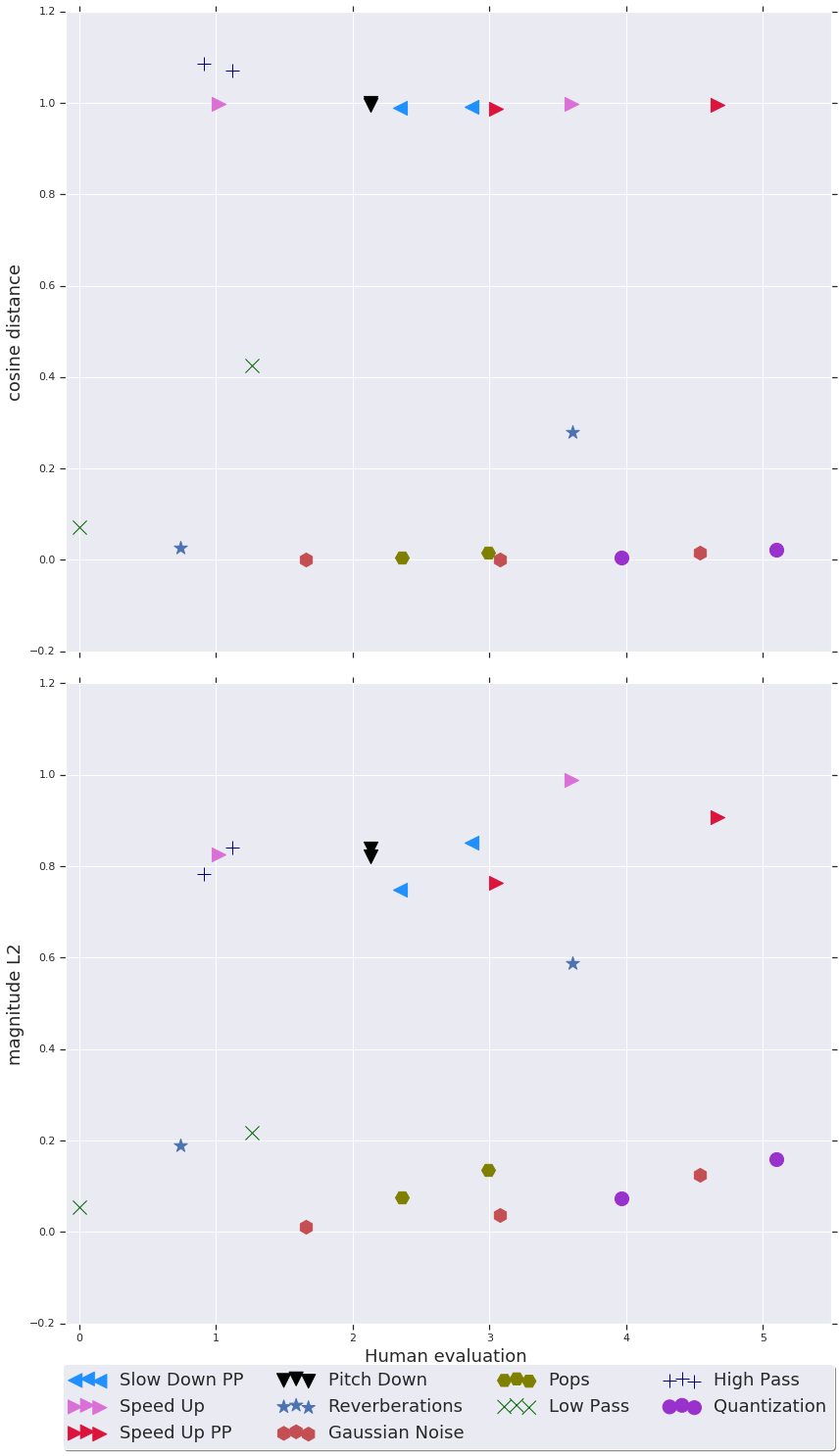}
\caption{\textit{Results of our human evaluation. The scale on x-axis is the worth value estimated by our Plackett-Luce model. The top plot compares this worth value to the distortions cosine distance and the bottom plot compares it to its magnitude L2 distance.}}
\label{fig:cos_human}
\end{figure}

Due to the time-consuming nature of the human evaluation, we only evaluated 10 distortions with total of $21$ parameter configurations on \num{300} audio segments (\SI{25}{minutes}) requiring \num{69300} pair-wise comparisons. After some training, the $20$ raters were able to compare and rate two \SI{5}{s} segments in under \SI{40}{s}.

The collected set of pair-wise evaluations was then ranked using a Plackett-Luce model \cite{plackettluce}, which estimates a \textit{worth value} for each parameter configuration. The evaluated distortions and their parameter configurations are listed in Appendix~\ref{app:params} together with their \gls{SDR} and \gls{FAD} scores.

Figure~\ref{fig:sdr_human} plots the worth values estimated by our Plackett-Luce model against both \gls{SDR} and \gls{FAD} scores. Neither of the plots shows a perfect correlation. \gls{SDR}, with a correlation coefficient of $0.39$, performs very poorly on \textit{speed up}, \textit{pitch preserving speed up/slow down}, \textit{reverberations} and \textit{pitch down} while correlating quite well with the other distortions.

The plot against \gls{FAD} also shows some outliers, most noticeably \textit{high pass} and \textit{low pass}. They are, however, still somewhat correlated and overall \gls{FAD}, with a correlation coefficient of $0.52$, correlates better than \gls{SDR} with how humans rate distortions. 

The other two examined metrics, cosine distance and magnitude L2 distance, are plotted against the human evaluation results in Figure~\ref{fig:cos_human}. With correlation coefficients of $-0.15$ for cosine distance and $-0.01$ for magnitude L2 distance, both perform significantly worse than either \gls{FAD} or \gls{SDR}. In particular, these two metrics fail at being able to compare between different types of distortions.

\section{Conclusion}
In this paper, we proposed the reference-free \gls{FAD} metric for measuring the quality of music enhancement approaches or models by comparing statistics of embeddings generated by their output to statistics of embeddings generated on a large set of clean music. Unlike other metrics, FAD can be computed using only a model's enhanced music output, without requiring access to either the original clean music or noise signal.

By testing a large, diverse set of artificial distortions, we show that \gls{FAD} can be useful in measuring the intensity of a given distortion. We compared it to traditional signal based evaluation metrics such as \gls{SDR}, and found that \gls{FAD} can be particularly useful for distortions which always lead to low \gls{SDR} scores independent of the distortion intensity. Our evaluation using human raters showed that \gls{FAD} correlated better with human ratings than \gls{SDR}. These results highlight the usefulness of \gls{FAD} as metric in measuring the quality of enhanced music and we hope to see others adopt it to report their results.

\section{Future Work}
Our goal was to develop a useful metric for evaluating music enhancement models and we have evaluated \gls{FAD} as such. However, we suspect that \gls{FAD} may also prove useful for evaluating a myriad of other audio enhancement and audio generation algorithms.

Although our evaluated set of distortions is quite large and diverse, it does not encompass all possible distortions that may occur to signals in either the real world or during enhancement. This is especially true if we wish to adapt \gls{FAD} to other audio domains. As an area of future work, we would like to evaluate the effectiveness of \gls{FAD} on further distortions and distortion combinations.

The VGGish embedding model uses log-mel features as input. Future work in this domain should investigate replacing VGGish with models that use other types of input such as raw samples or a complex spectrogram. A key disadvantage of our implementation of \gls{FAD} is that it only looks at embeddings created on \SI{1}{s} windows, which means that the metric is unaware of long distance temporal changes within a song. An embedding model which operates on music of variable lengths and computes a single embedding per song may be useful here.

\section{Acknowledgements}
The authors would like to thank Javier Cabero Guerra, Pierre Petronin, Trisha Sharma and all the participating raters for helping to conduct our large-scale human evaluation. We thank Kevin Wilson for the insightful comments that have greatly improved this publication. We further thank Marvin Ritter, F\'elix de Chaumont Quitry, Dan Ellis, Dick Lyon, Sammy El Ghazzal, David Ramsay, and the Google Brain Z\"urich team for their support and helpful conversations.


\nocite{unterthiner2018towards}

\clearpage
\printbibliography

\clearpage

\appendix

\section{Window Step Size}
\label{app:window_step}
The VGGish embedding model that we are using requires an input of \SI{1}{s} of audio. When extracting embeddings from a continuous stream of audio we can either partition the stream into \SI{1}{s} long chunks or extract embeddings from a \SI{1}{s} moving window every $t$ seconds. Using a small embedding window step length $t$ will provides us with more embeddings, which may result in us being able to estimate the multivariate Gaussians better.

\begin{figure}[h]
\includegraphics[width=1\linewidth]{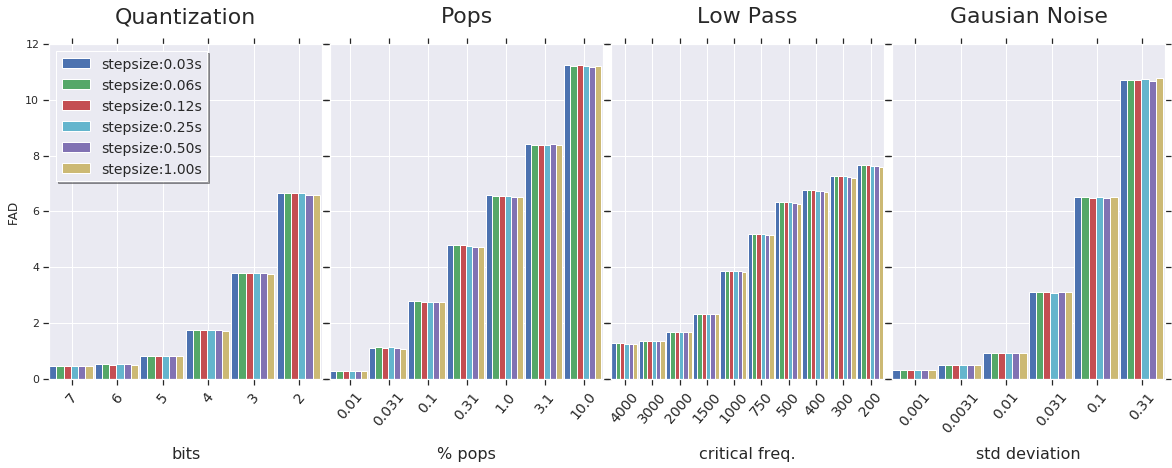}
\caption{\textit{Overview of the effect of the embedding window step length on a subset of the evaluated distortions.}}
\label{fig:step}
\end{figure}

We compare various embedding window step lengths to determine whether smaller values are useful. The results on some of our distortion configurations can be seen in Figure~\ref{fig:step}. Overall the \gls{FAD} scores change very little as we reduce the embedding window step length, indicating that computing many embeddings from highly overlapping segments is not necessary. For a couple of distortion types having non-overlapping windows does change the \gls{FAD} score slightly, and we therefore recommend using an embedding window step length of \SI{0.5}{s}.

\clearpage

\section{Evaluation Set Size}

As \gls{FAD} requires measuring the distance between two multivariate Gaussians estimated on sets of embeddings, it can be greatly affected by the size of these sets. Using smaller sets will result in a less accurate estimate of the multivariate Gaussians. In our case, we assume that our set of background embeddings is significantly larger than the set of evaluation embeddings and investigate how large this set needs to be in order to have a stable \gls{FAD} score.

As described in Section~\ref{sec:data}, we split our evaluation data into \SI{5}{s} long audio clips. We are able to extract around \num{40000} of these  \SI{5}{s} long audio clips from our full evaluation set. We apply distortions with various parameter combinations and extract embeddings using an embedding window step length of \SI{0.5}{s}.

As possible evaluation set sizes, we consider k audio clips, with k being either \num{100}, \num{200}, \num{300}, \num{400}, \num{500}, \num{1000}, \num{5000} or \num{10000}. For each of these possible sizes, we compute the \gls{FAD} scores of our distortions at various parameter combinations multiple times using different subsets of evaluation audio clips, allowing us to examine how the evaluation set size affects the variance in \gls{FAD}.

Different distortion types and configurations will have different expected variances, e.g. \textit{Gaussian noise} with a stddev of \num{0.1} and \textit{speed up} of \SI{1}{\%}. To compensate for this, we therefore we compute the index of dispersion $D = \frac{\sigma^2}{\mu}$ for each distortion configuration which normalizes the variances by the mean.

\begin{figure}[h]
\includegraphics[width=0.3\linewidth]{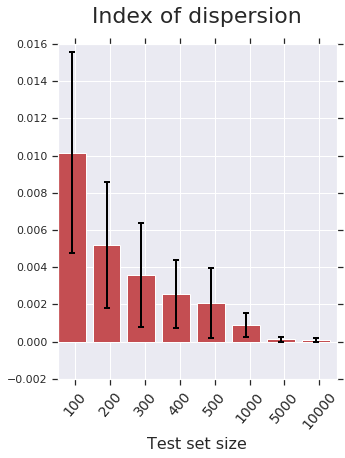}
\caption{\textit{Average index of dispersion for the test evaluation set sizes across all examined distortion parameter configurations.}}
\label{fig:index}
\end{figure}

The average index of dispersion across all distortion configuration can be seen in Figure~\ref{fig:index}. The very high value for \num{100} indicates that \num{100} audio clips or 8 minutes and 20 seconds of audio is not enough to compute a stable \gls{FAD} score. An ideal amount would be about 5000 audio clips or around 7 hours. This is a lot of data for evaluation purposes and will often not be available. While not as stable as larger evaluation set sizes, we begin to get usable results from about \num{300} audio clips or 25 minutes of audio.

\clearpage

\section{Evaluated Distortion Parameter Configurations}
\label{app:params}

\definecolor{LightCyan}{rgb}{0.88,1,1}

\begin{table}[ht]
\centering
  \begin{tabular}{|p{3cm}|p{2.5cm}|c|}
  \hline
Distortion & Parameter & Parameter values \\ \hline
Slow Down & factor & 1.01, 1.02, 1.05, 1.1, 1.2, 1.3, 1.5, 1.7, 2, 2.5, 3, 4, 5 \\
\rowcolor{LightCyan}
Slow Down PP & factor & 1.01, 1.02, 1.05, 1.1, 1.2, 1.3, 1.5, 1.7, 2, 2.5, 3, 4, 5 \\
Speed Up & factor & 0.99, 0.98, 0.95, 0.9, 0.8, 0.7, 0.6, 0.5, 0.4, 0.2, 0.1 \\

\rowcolor{LightCyan}
Speed Up PP & factor & 0.99, 0.98, 0.95, 0.9, 0.8, 0.7, 0.6, 0.5, 0.4, 0.2, 0.1 \\
Pitch Up & semi-tone & 0.05, 0.1, 0.15, 0.2, 0.25, 0.5, 0.75, 1, 1.5, 2, 2.5, 3, 4, 5 \\

\rowcolor{LightCyan}
Pitch Down & semi-tone & 0.05, 0.1, 0.15, 0.2, 0.25, 0.5, 0.75, 1, 1.5, 2, 2.5, 3, 4, 5 \\
Reverberations & dampening \newline delay: \SI{1}{s}\newline echos: 3 & 0.1, 0.2, 0.3, 0.4, 0.5, 0.6, 0.7, 0.8, 0.9 \\

\rowcolor{LightCyan}
Reverberations & dampening \newline delay: \SI{0.5}{s}\newline echos: 3& 0.1, 0.2, 0.3, 0.4, 0.5, 0.6, 0.7, 0.8, 0.9 \\
Reverberations & dampening \newline delay: \SI{0.25}{s}\newline echos: 3& 0.1, 0.2, 0.3, 0.4, 0.5, 0.6, 0.7, 0.8, 0.9 \\

\rowcolor{LightCyan}
Reverberations & dampening \newline delay: \SI{0.25}{s}\newline echos: 5& 0.1, 0.2, 0.3, 0.4, 0.5, 0.6, 0.7, 0.8, 0.9 \\
Gaussian Noise & std deviation & 0.0001, 0.00031, 0.001, 0.0031, 0.01, 0.031, 0.1, 0.31 \\

\rowcolor{LightCyan}
Pops & \% pops & 0.0001, 0.00031, 0.001, 0.0031, 0.01, 0.031, 0.1, 0.31 \\
Low Pass & critical freq. & 4000, 3000, 2000, 1500, 1000, 750, 500, 400, 300 \\

\rowcolor{LightCyan}
High Pass & critical freq. & 200, 300, 400, 500, 750, 1000, 1500, 2000, 3000, 4000\\
Quantization & bits & 9, 8, 7, 6, 5, 4, 3, 2 \\

\rowcolor{LightCyan}
Griffin Lim & iterations & 500, 200, 100, 50, 20, 10, 5, 1 \\
Griffin Lim Zero & iterations & 500, 200, 100, 50, 20, 10, 5, 1 \\

\rowcolor{LightCyan}
Mel Filter Wide & num. bands & 264, 128, 64, 32 \\
Mel Filter Narrow & num. bands & 264, 128, 64, 32, 16, 8 \\ \hline
  \end{tabular}
    \caption{\textit{List of all the examined distortions and their parameter configurations.}}
    \label{table:params}
\end{table}
    
\begin{table}[ht]
  \centering
  \begin{tabular}{|p{4cm}|p{5cm}|p{1.5cm}|p{1cm}|p{1cm}|}
  \hline
Distortion & Parameters & Worth & FAD  & SDR  \\ \hline
low pass &critical frequency: 5000 & -0.00  & 0.94 & 56 \\
\rowcolor{LightCyan}
reverberations & dampening: 0.2\newline delay: \SI{1}{s}\newline echos: 3 & -0.74   & 0.24 & 13 \\
high pass & critical frequency: 400 & -0.92   & 1.46 & 41 \\
\rowcolor{LightCyan}
speed up & factor: 0.95 & -1.02   & 0.19 & -21 \\
high pass & critical frequency: 500 & -1.12  & 2.34 & 41 \\
\rowcolor{LightCyan}
low pass & critical frequency: 1500 & -1.26   & 2.39 & 48 \\
added gaussian noise & stddev: 0.0031  & -1.66 & 0.55 & 36\\
\rowcolor{LightCyan}
pitch down & semi-tone: 0.25 & -2.13  & 0.63 & -21 \\
pitch down & semi-tone: 0.1 & -2.13   & 0.65 & -21 \\
\rowcolor{LightCyan}
slow down pp & factor: 1.05 & -2.35   & 2.25 & -3 \\ 
pops & percentage: 0.00031 & -2.36   & 1.24 & 21 \\
\rowcolor{LightCyan}
slow down pp & factor: 1.2 & -2.87   & 3.37 & -10\\
pops & percentage \%: 0.001 & -2.99   & 2.80 & 16 \\
\rowcolor{LightCyan}
added gaussian noise & stddev 0.01 & -3.00   & 0.94 & 26 \\
speed up pp & factor: 0.95 & -3.05   & 1.43 & -5 \\ 
\rowcolor{LightCyan}
speed up & factor: 0.8 & -3.60   & 0.82 & -21 \\
reverberations & dampening: 0.4\newline delay: \SI{0.25}{s}\newline echos: 5& -3.61  & 1.08 & 1 \\
\rowcolor{LightCyan}
quantization & bits: 4 & -4.00  & 1.63 & 20 \\
added gaussian noise & stddev 0.031 & -4.54  & 2.93 & 16 \\
\rowcolor{LightCyan}
speed up pp & factor: 0.8 & -4.67  & 2.58 & -12 \\
quantization  & bits: 3 & -5.10  &  3.50 & 14 \\ \hline
  \end{tabular}
    \caption{\textit{List of human evaluated distortions, their parameter configurations, worth value estimated by our Plackett-Luce model and both their \gls{SDR} and \gls{FAD} scores.}}
    \label{table:results}
\end{table}

\end{document}